\documentstyle[12pt,subeqn,cite]{article}

\input epsf.tex
\def\DESepsf(#1 width #2){\epsfxsize=#2 \epsfbox{#1}}

\def\etal{ {\em et al.}}

\def\be {\begin{equation}}
\def\ee {\end{equation}}
\def\barr{\begin{array}}
\def\earr{\end{array}}
\def\dis{\displaystyle}
\def\ra{\rightarrow}

\def\bra {\langle}
\def\ket{\rangle}

\def\l {\lambda}
\def\rp {R_p}
\def\rpv {R_p\!\!\!\!\!\!/~~}
\def\hlp {H_{eff}^{\lambda'}}
\def\hlpp {H_{eff}^{\lambda''}}
\def\betapk {B\rightarrow \eta' K}
\def\bpletapk {B^{\pm}\rightarrow \eta' K^{\pm}}
\def\betak {B\rightarrow \eta K}
\def\bpletak {B^{\pm}\rightarrow \eta K^{\pm}}
\def\msnu {m_{\tilde\nu_{iL}}}
\def\msnus {m_{\tilde\nu_{iL}}^2}
\def\msells {m_{\tilde e_{iL}}^2}

\def\msurs {m_{\tilde  u_{iR}}^2}

\def\rtwo {\sqrt{2}}
\def\lappeq{\mathrel{\rlap{\raise.5ex\hbox{$<$}}
                    {\lower.5ex\hbox{$\sim$}}}}
\def\Black{}
\def\Blue{}

\def\RawSienna{}
\renewcommand{\thefootnote}{\fnsymbol{footnote}}
\begin{document}
\begin{flushright}  MRI-PHY/P981064\\CTP-TAMU-47-98\\   {\large \tt hep-ph/9812209}
\end{flushright}

\vskip 2 true cm

\begin{center}

{\Large\bf \Blue 
A supersymmetric resolution of the anomaly in charmless nonleptonic 
{\boldmath $B$}-decays}\\[0.5cm] 
\Black 
\large\em  Debajyoti Choudhury$^{a,}$\footnote{Electronic address: 
debchou@mri.ernet.in},   
B. Dutta$^{b,}$\footnote{Electronic address:  
b-dutta@rainbow.physics.tamu.edu},  
Anirban Kundu$^{a,c}$\footnote{Electronic address: akundu@mri.ernet.in}
\\[1cm]
${ }^a$ Mehta Research Institute, Chhatnag Road, Jhusi, Allahabad - 211 019, 
India\\
${ }^b$ Center for Theoretical Physics, Department of Physics, Texas A \& M University, College Station, TX 77843, USA\\
${ }^c$ Department of Physics, 
                              Jadavpur University, Calcutta - 700 032, India

\end{center}

\vskip 2 true cm

\RawSienna
\begin{abstract}

We examine the  large branching ratio for the process $B\ra \eta' K$ 
from the standpoint of R parity violating supersymmetry. We have given all
possible $\rpv$ contributions to 
  \Blue
  $B\ra \eta' K$ 
\RawSienna
amplitudes. We find that 
only  two pairs of $\l'$-type $\rpv$ couplings can solve this problem 
after satisfying all other experimental bounds.  We also analyze those 
modes where these couplings can appear, {\em e.g.}, $B^\pm \ra \pi^{\pm}K^0$, 
$B^{\pm,0} \ra K^{*\pm,0} \eta^{(\prime)}$, $B^{\pm} \ra \phi K^{\pm}$
etc., and predict their branching ratios.   Further, one of these 
two pairs of couplings is found to lower the branching  ratio of
$B^{\pm}\ra \phi K^\pm$, thereby allowing larger 
$\xi\equiv{1\over N_c}$. This allows us to fit $B^{\pm}\ra \omega K^{\pm}$  and
$B^{\pm}\ra \omega \pi^{\pm}$, which could not be done in the SM framework.

\end{abstract}

\Black
\clearpage
\setcounter{page}{1}
\pagestyle{plain}
\setcounter{footnote}{0}
\renewcommand{\thefootnote}{\arabic{footnote}}
\centerline{\bf I. Introduction}

Recently, the CLEO collaboration has reported the branching ratios (BR) of 
a number of charmless nonleptonic $B\ra PP$ and $B\ra PV$ two-body decay modes
where $P$ and $V$ denote, respectively, a pseudoscalar and a vector meson. 
Some of these modes have been observed for the first time and the  
upper bounds on the others have been improved\cite{cleo_talks,cleo_papers}.

Among the $B\ra PP$ modes,  the branching ratio for  $\bpletapk$  
is found to be larger 
than that expected  within the Standard Model (SM). 
This result has initated lots of investigations in the last one year
\cite{Kag_Pet, Dat_He_Pak, Ali_Greub, Cheng_Tseng}. 
 This kind of unexplained 
puzzle also exists in the $B\ra PV$
modes where it is found that the  branching ratios of $B^{\pm}\ra  
\phi K^{\pm}$, $ B^{\pm}\ra \omega  \pi^{\pm}$ and $ B^{\pm}\ra \omega 
K^{\pm}$ are hard to fit simultaneously \cite{desh1,ali1}.  
Present attempts to explain the large branching ratio
$BR(B^{\pm} \ra \eta' K^{\pm})$ involve  large form factors and/or large  charm
content for $\eta'$, with contribution arising from
$b\ra s\bar c c\ra s\eta'(\eta)$, and low strange quark 
mass \cite{Kag_Pet, Dat_He_Pak, Ali_Greub, Cheng_Tseng}. 
In an interesting paper \cite{ros}, consequences of large
$\betapk$ branching ratio from purely SU(3) viewpoint has been studied. 

In this paper we try to address the large BR problem from the standpoint of
$R$-parity ($\rp$) violating supersymmetry (SUSY) theories.  
Motivations for invoking
SUSY and its $\rpv$ version have been discussed  in detail in the literature
\cite{susy}. 
Some of its effects on $B$-decays have also been investigated \cite{guetta}.
Since the new interactions modify 
the SM Hamiltonian,
it is  natural to revisit these calculations and try to see
whether the above mentioned puzzles can be solved.  
We calculate the QCD-improved
short-distance part with the  usual operator product expansion and Wilson
coefficients (WC), while  the long-distance parts are calculated  by the
factorization technique which is very successful in estimating $B\ra D$ decays.
The requirement that any ``new physics" solution of the perceived anomaly 
does not overly affect other observables that are in good agreement with the 
SM predictions
restricts us to two particular sets of couplings within the $\rpv$ scenario. 
Interestingly enough,  we find that one of these sets also leads  
to a better fit for the decays  $B^{\pm}\ra \phi K^{\pm}$, 
$B^{\pm}\ra \omega \pi^{\pm}$ and 
$B^{\pm}\ra \omega K^{\pm}$.  

We organize the letter as follows. 
In section II, we give a very brief introduction
to the SM and $\rpv$ Hamiltonian, and list the  possible $\rpv$ 
operators  that can contribute to
charmless decays. We 
discuss the $B\ra PP$ and $B\ra PV$ decay modes 
in section III. The new physics contributions to the decay modes 
$\bpletapk$ and $\bpletak$ are shown. 
In section IV, we discuss  how $\rpv$ can  
raise the branching ratio of $B\ra \eta'K$ without jeopardizing 
other decay modes. We make predictions about the
yet-to-be-observed channels which can be tested in the 
upcoming B-factories. We also
discuss how to fit the new results in $B\ra PV$ modes  in presence of the new
couplings  which are used to raise the BR of
$B\ra \eta' K$. We conclude in section V.

\newpage    

\centerline{\bf 2. Effective Hamiltonian for charmless decays } 
\centerline{\bf 2.1 SM Hamiltonian}

The effective Hamiltonian for charmless nonleptonic $B$ decays can be 
written as
\be {\cal H} = {G_F\over\rtwo}\Big[ V_{ub}V^*_{uq}\sum_{i=1,2} c_iO_i
   - V_{tb}V^*_{tq}\sum_{i=3}^{12} c_iO_i\Big] + h.c.
\ee
 The Wilson coefficients (WC), $c_i$, take care of the short-distance QCD
corrections. We find all our expressions in terms of the effective WCs and  
refer the reader to  the papers \cite{desh,buras,ciuchini,kramer}  
for a detailed 
discussion\footnote{Since the $\rpv$ operators will be shown to be 
         small, their mixing with the SM operators may safely be 
         neglected at the current level of accuracy.}. 
We use the effective WCs for the processes
$b\ra s\bar qq'$ and $b\ra d\bar qq'$ from ref.~\cite{desh}. 
The regularization scale is taken to be $\mu=m_b$.  
In our subsequent discussion, we will neglect small effects 
of the  electromagnetic moment operator
$O_{12}$, but will take into account  effects from the four-fermion operators
$O_1-O_{10}$ as well as the chromomagnetic operator $O_{11}$.

\vskip 0.3 true cm
\centerline{\bf 2.2 The $R_p$-violating Hamiltonian }

The superpotential of the minimal supersymmetric standard model 
(MSSM) can contain terms, apart from those obtained by a straightforward  
supersymmetrization of the SM potential, of the form
\be 
 {\cal W}_{\rpv}=\kappa_iL_iH_2 + \l_{ijk}L_iL_jE_k^c + \l'_{ijk}L_iQ_jD_k^c
          + \l''_{ijk}U_i^cD_j^cD_k^c \, 
      \label{superpot}
\ee  
where $E_i$, $U_i$ and $D_i$ are respectively the $i$-th type of  lepton,
up-quark and down-quark singlet superfields, $L_i$ and
$Q_i$ are the SU$(2)_L$ doublet lepton and quark superfields, and
$H_2$ is the Higgs doublet with the appropriate hypercharge.  
Symmetry properties dictate that $\l_{ijk}=-\l_{jik}$ and 
$\l''_{ijk}=-\l''_{ikj}$. Apparently,
the bilinear term  can be rotated away with a redefinition of lepton and  Higgs
superfields, but the effect reappears as $\l$s, $\l'$s and 
lepton-number violating soft terms \cite{roybabu}. 
The first three  terms of eq.(\ref{superpot}) violate
lepton number  whereas the fourth term violates baryon number.  
Thus, simultaneous presence of both sets 
would lead to catastrophic rates for proton
decay, and hence it is tempting to invoke a discrete symmetry which  
forbids all such terms. One introduces the conserved quantum number
\[
\rp = (-1)^{3B+L+2S}
\]  which is $+1$ for the SM particles and $-1$ for their superpartners.  
However, to prevent proton decay, one needs to forbid only one set,  
and not necessarily  both.  This leaves us with the possibility  
of additional Yukawa interactions within the
MSSM, many consequences of which have already been discussed extensively 
in the literature.
  
For our purpose, we will assume either $\l'$ or $\l''$-type 
couplings to be present
($\l$-type couplings do not lead to nonleptonic decays), 
but not both. Assuming all
$\rpv$ couplings to be real, the effective Hamiltonian for charmless nonleptonic
$B$-decay can be written  as\footnote{In this paper, we will not consider the
CP-violating effects
   of these couplings, {\em i.e.}, we will assume all of them to be real. 
   However, the fact that they may not all be real leads to 
   interesting consequences.}

\subequations
\be
\barr{rcl}
\dis \hlp (b\ra \bar d_j d_k d_n)
   & = & \dis d^R_{jkn} [ \bar d_{n\alpha} \gamma^\mu_L d_{j\beta}
                           \; \bar d_{k\beta} \gamma_{\mu R} b_{\alpha}] 
            + d^L_{jkn} [ \bar d_{n\alpha} \gamma^\mu_L b_{\beta}
                           \; \bar d_{k\beta} \gamma_{\mu R} d_{j\alpha}] 
              \ ,
             \\[1.5ex]
\dis \hlp (b\ra \bar u_j u_k d_n)
   & = & \dis u^R_{jkn} [\bar u_{k\alpha} \gamma^\mu_L u_{j\beta}
                           \; \bar d_{n\beta} \gamma_{\mu R} b_{\alpha}]
              \ ,
              \\[1,5ex]
\dis \hlpp (b\ra \bar d_j d_k d_n)
   & = &  \dis {1\over 2} d''_{jkn}
            [ \bar d_{k\alpha} \gamma^\mu_R d_{j\beta}
                \; \bar d_{n\beta} \gamma_{\mu R} b_{\alpha} 
            - \bar d_{k\alpha} \gamma^\mu_R d_{j\alpha}
                \; \bar d_{n\beta} \gamma_{\mu R} b_{\beta} ]
           \ ,
              \\[1.5ex]
\dis \hlpp (b\ra \bar u_j d_k d_n)
        & = &  \dis 
            u''_{jkn} [ \bar u_{k\alpha} \gamma^\mu_R u_{j\beta}
                         \; \bar d_{n\beta} \gamma_{\mu R} b_{\alpha} 
                       - \bar u_{k\alpha} \gamma^\mu_R u_{j\alpha}
                         \; \bar d_{n\beta}\gamma_{\mu R}b_{\beta}]
           \ ,
\earr
    \label{rp_hamilt}
\ee   
with
\be
 \barr{rclcrclcl} d^R_{jkn} &=& \dis 
      \sum_{i=1}^3 {\l'_{ijk}\l'_{in3}\over 8\msnus},
      &  &  d^L_{jkn} &=& \dis \sum_{i=1}^3 {\l'_{i3k}\l'_{inj}\over 8\msnus},
      &  & (j,k,n=1,2)
           \\[1.5ex]  u^R_{jkn} &=& \dis \sum_{i=1}^3 {\l'_{ijn}\l'_{ik3}\over
8\msells},
      &  & 
      &  & 
      &  & (j,k=1, \ n=2)
          \\[1.5ex] d''_{jkn} & = & \dis \sum_{i=1}^3 {\l''_{ij3}\l''_{ikn}\over
4\msurs},
      &  &  u''_{jkn} &=& \dis \sum_{i=1}^2 {\l''_{ji3}\l''_{kin}\over 4\msurs},
      &  & (j=1,2,~k=1,~n=2).
\earr
\ee
\endsubequations 
where $\alpha$ and $\beta$ are colour indices and 
$\gamma^\mu_{R, L} \equiv \gamma^\mu (1 \pm \gamma_5)$. 
The  parenthetical remarks on
the subscripts concentrate on  only the relevant couplings.

As is obvious, data on low energy processes can be used to impose rather strict
constraints on many of these  couplings~\cite{constraints,products,herbi}. 
Most such
bounds have been  calculated under the assumption of there being 
only one  non-zero
$\rpv$ coupling. From eq.(\ref{rp_hamilt}), it  
is evident that such a supposition
precludes any tree-level flavour-changing neutral 
currents, thus negating the very aim
of this paper. However, there is no strong argument 
in support of only one $\rpv$  coupling being nonzero. In fact, 
it might be  argued~\cite{products} that a hierarchy of
couplings may be  naturally obtained on account of the mixings in either of the 
quark and squark sectors. In this paper we will take a more 
phenomenological approach and try to find out the values of 
such $\rpv$ couplings for which  all available data
are satisfied. An important role will be played by  
the $\l'_{32i}$ type couplings, the constraints on which are relatively weak.

\vskip 1 true cm
\centerline{\bf 3. \boldmath{$B\ra PP$} and \boldmath{$PV$} modes}

We consider next the  matrix elements of the various vector 
($V_\mu$) and  axial vector ($A_\mu$) quark currents between 
generic meson states. For the decay constants
of a pseudoscalar ($P$) or a vector ($V$) meson defined through
\subequations
\be
\barr{rcl}
\bra 0|A_\mu|P(p)\ket & = & \dis if_P p_\mu \\
\langle 0|V_{\mu}|V(\epsilon,p)\rangle 
                      & = & f_{V} m_{V}\epsilon_{\mu} \ ,
\earr
\ee 
we use the following (all values in MeV)~\cite{Ali_Greub}, 
\be  
  f_\omega = 195,\ f_{K^*}  = 214,\ f_\rho = 210,\ f_\pi = 134,\ f_K  =
158,\ f_{\eta_1} = 1.10f_\pi,\ f_{\eta_8} = 1.34 f_\pi.
\ee  
The decay constants of the mass eigenstates  $\eta$ and $\eta'$ are related to
those for the weak eigenstates through the relations
\[
\barr{rclcrcl} f^u_{\eta^{\prime}} & = & \dis {f_8\over\sqrt{6}} \sin \theta 
                     +{f_1\over\sqrt{3}} \cos \theta
     & \qquad & 
    f^s_{\eta^{\prime}} & = & \dis -2{f_8\over\sqrt{6}} \sin \theta
                              +{f_1\over\sqrt{3}} \cos \theta
      \\ f^u_{\eta} & = & \dis {f_8\over\sqrt{6}} \cos \theta 
                - {f_1\over\sqrt{3}} \sin \theta, 
     & \qquad & 
  f^s_{\eta} & = & \dis -2{f_8\over\sqrt{6}} \cos \theta
                            -{f_1\over\sqrt{3}} \sin \theta.
\earr
\]
\endsubequations 
The mixing angle can be inferred from the  data on the
$\gamma\gamma$ decay modes\cite{angle} to be
$\theta \approx -22^\circ$.

Whereas the only nonzero $B\ra P$  matrix element can be parametrized as
\subequations
\be
\bra P(p')|V_\mu|B(p)\ket = \Big[ (p'+p)_\mu - {m_B^2-m_P^2\over q^2}q_\mu\Big ]
   F_1^{B\ra P} 
 + {m_B^2-m_P^2\over q^2}q_\mu F_0^{B\ra P} \ ,
\ee  
the $B\ra V$ transition is given by
\be
\barr{l}
\langle V(\epsilon, p')|(V_{\mu}-A_{\mu})|B(p)\rangle
    \\
\dis  \hspace*{1em} =  
  {2 V \over {m_B+m_V}} 
     \epsilon_{\mu\nu\alpha\beta}\epsilon^{*\nu} p^{\alpha} p^{\prime\beta}
         \\ 
\dis  \hspace*{1em} +  
    i \left[ 
           (m_B+m_V) A_1 \: \epsilon^*_\mu + 
               \epsilon^* \cdot q 
                  \: \left\{- A_2 \frac{(p+p')_\mu}{m_B+m_V} 
                        + 2 m_V \frac{q_\mu}{q^2} \left( A_0 - A_3 
                                            \right)
                  \right\}
                  \right]
\earr
\ee 
with $2m_VA_3 \equiv (m_B + m_V) A_1 - (m_B - m_V) A_2$.  
All of the quantities $F_{0,1}^{B \ra P}$, 
$V^{B \ra V}$ and $F_{0,1}^{B \ra V}$ have a formfactor behaviour in $q^2 
\equiv (p - p')^2$. 
Note that $F_1=F_0$ at $q^2=0$, and, to a very good approximation,  we can set
$F(m_{P_2}^2) = F(0)$  for $B$ decay formfactors since the $q^2$ dependence  is
dominated by meson poles at the scale $m_B$.  
Flavour $SU(3)$ then allows us to write 
\be
\barr{rclcrcl} F^{B\ra K, \pi^\pm}_{0,1} & = & F \ , 
        & \qquad & 
     F^{B\ra \pi^0}_{0,1} & = & \dis {F\over\rtwo} \ ,  
             \\   F^{B\ra \eta'}_{0,1} & = & \dis 
            F \left({\sin\theta\over\sqrt{6}} + {\cos\theta\over\sqrt{3}}
              \right) \ ,  
        & \qquad &
     F^{B\ra \eta}_{0,1} & = & \dis 
            F \left({\cos\theta\over\sqrt{6}} - {\sin\theta\over\sqrt{3}}
              \right) 
          \ .
   \label{fs}
\earr
\ee  
There seems to be considerable variation in the range of $F$ estimated in the
literature. 
Bauer {\em et al} estimate it to be $0.33$ \cite{stech} while Deandrea
{\em et al} get a value of $0.5$ \cite{gatto}.  We find that while within the 
SM,
the combination ($F=0.36, |V_{ub}/V_{cb}| = 0.07$) yields
a good fit to $B\ra \pi\pi$ and $B\ra \pi K$ data \cite{desh}, 
introduction of $\rpv$ interactions allows larger values of $F$.
As for the $B \ra V$ formfactors, it can easily 
be ascertained that,  of the four, only $A_0$ is relevant
for  the $B \ra P V$ decays that we are interested in.  For the current $\bar
u\gamma_{\mu}(1-\gamma_5)b$,  we have
\be   A_0^{B\ra \omega}={G\over \sqrt2},
        \qquad   A_0^{B\ra K^*}=G ,
        \qquad   A_0^{B\ra\rho}={G\over \sqrt 2},
\ee 
\endsubequations 
where we use G=0.28~\cite{Ali_Greub}. The only remaining parameters of
interest is the mass of the strange quark  
for which we use $m_s(1~{\rm GeV}) = 165$ MeV leading to
$m_s(m_b) = 118$ MeV.

\vskip 0.3 true cm
\centerline{\bf 3.1 \boldmath{$B^\pm\ra \eta'(\eta) K^\pm$}}

The effective SM  Hamiltonian for this decay and its matrix elements  are
well-studied and can be found in Refs.\cite{Ali_Greub,desh}. As for the 
$\rpv$ operators, it is easy to see that only six 
of them  may contribute (with none from the $u''$ set) and may 
be expressed in  terms of 
\[ 
\barr{rcl} A_{M_1} & = & \bra M_2|J_b^\mu|B\ket \; \bra M_1|J_{l\mu}|0\ket\\  A_{M_2}
& = &\bra M_1|J_b^{'\mu}|B\ket \; \bra M_2|J'_{l\mu}|0\ket. 
\earr
\] 
where $J$ and $J'$ stand for quark currents and the subscripts $b$ and 
$l$ indicate whether the current involves a $b$ quark or only the light quarks.
Neglecting the annihilation  diagrams\footnote{Such processes cannot be
         treated under the factorization ansatz, but are expected 
         to be negligibly small in any case.} we have, for the $\betak$  matrix
elements,
\subequations 
\be
\barr{rcl} 
{\cal M}^{\l'} &=& \dis
       \left( d^R_{121} - d^L_{112} \right) \xi A_\eta^u 
        +  
          \left( d^L_{222} - d^R_{222} \right) \: 
                 \left[  \frac{\bar m}{m_s}
                   \left(A_\eta^s -A_\eta^u\right)- \xi  A_\eta^s\right]
        \\[1.5ex]
       & + &  \dis 
        \left(d^L_{121} - d^R_{112} \right) \frac{\bar m}{m_d} A_\eta^u
       + 
       u^R_{112} \left[\xi A_\eta^u - {2 m_K^2 A_K \over (m_s+m_u) (m_b-m_u)} 
                 \right],
\earr
\ee 
where $\bar m \equiv m^2_\eta / (m_b - m_s)$, 
and\footnote{Note that $\bra 0|\bar s i \gamma_5 s|\eta^{(\prime)}\ket =
             -(f^s_{\eta^{(\prime)}}-
             f^u_{\eta^{(\prime)}})m^2_{\eta^{(\prime)}}/ 2 m_s$~\protect
             \cite{Ali_Greub}.}  
\be {\cal M}^{\l''} = d''_{112}(1-\xi)A_{\eta}^u.
\ee 
\endsubequations
Analogous expressions hold for $\bpletapk$ where  we have to replace
$A^u_{\eta}$ by $A^u_{\eta^{\prime}}$, 
  $A^s_{\eta}$ by $A^s_{\eta^{\prime}}$ and $m_{\eta}$ by 
  $m_{\eta^{\prime}}$. We note that
$\l''_{112}$ and $\l''_{113}$ are bounded to be very small  irrespective of the
presence of other $\rpv$ operators, and hence may be neglected. 
For the numerical analysis, we
take $m_{\eta_8} =m_{\eta}$ and $m_{\eta_1}=m_{\eta'}$.

\vskip 1 true cm
\begin{table}
\begin{center}
\begin{tabular}{|| l | l | l ||   l | l | l ||}
\hline
\multicolumn{1}{|| c | } {Mode }
    & \multicolumn{1}{ c | } {$BR \times 10^5 $} 
    & \multicolumn{1}{ c || } {SM theory $\times 10^5$}
       & \multicolumn{1}{ c | } {Mode }
              & \multicolumn{1}{ c | } {$BR \times 10^5 $} 
              & \multicolumn{1}{ c || } {SM theory $\times 10^5$}
       \\
\hline
$B^+ \ra \eta' K^+$  & $6.5^{+1.5}_{-1.4} \pm 0.9$ & $0.8-4.3$   
       & $B^0 \ra \eta' K^0$  & $4.7^{+2.7}_{-2.0} \pm 0.9$
       & $0.7-4.1$
          \\
$B^+ \ra \eta' K^{\ast+}$  & $< 13  $ & $0.01-0.18$
       & $B^0 \ra \eta' K^{\ast 0}$  & $< 3.9$ & $0.03-0.18$
          \\
$B^+ \ra \eta K^+$  & $ < 1.4 $  & $0.06-0.14$
       & $B^0 \ra \eta K^0$  & $ < 3.3 $ & $0.03-0.14$ 
         \\
$B^+ \ra \eta K^{\ast +}$  & $ < 3.0 $ & $0.14-0.31$ 
       & $B^0 \ra \eta K^{\ast 0}$  & $ < 3.0 $ & $0.1-0.5$ 
         \\
\hline
$B^+ \ra \pi^+ K^0$  & $ 2.3^{+1.1}_{-1.0} \pm 0.4 $& $1.1-3.5$ 
       &  $B^0 \ra \pi^0 K^0$  & $ < 4.1 $& $0.6-1.9$ 
          \\
$B^+ \ra \pi^0 K^+$  & $ < 1.6 $ & $1.0-1.4$ 
       &  $B^0 \ra \pi^- K^+$  & $ 1.5^{+0.5}_{-0.4}\pm 0.1$ & $ 1.1-2.1$ 
          \\
$B^+ \ra \pi^+ \pi^0$  & $ <2.0 $ & $ 0.3-1.3$ 
       &  $B^0 \ra \pi^+ \pi^-$  & $ <1.5$ & $0.8-1.5$ 
          \\
\hline
$B^+ \ra \phi K^+$  & $ < 0.53 $ & $ 0.07-5.0 $ 
       &    &  & 
          \\
$B^+ \ra \omega K^+$  & $ 1.5^{+0.7}_{-0.4} \pm 0.3 $ & $ 0.01-3.5$
       
       & $B^+ \ra \omega \pi^+$  & $ 1.1^{+0.6}_{-0.5} \pm 0.2 $ & $ 0.06-1.7$ 
          \\
\hline
\end{tabular}
\end{center}
\caption{\em Branching ratios (or upper bounds) for various $B$-meson 
	 decays. Also shown are the theoretical predictions based on
	 the SM only \protect\cite{sm_branch}. }
	\label{tab:1}
\end{table}

\centerline{\bf 4. Analysis} 

\begin{figure}[htb]
\hspace*{-0.5cm}
\centerline{
\epsfxsize=6.2cm\epsfysize=7.0cm
\epsfbox{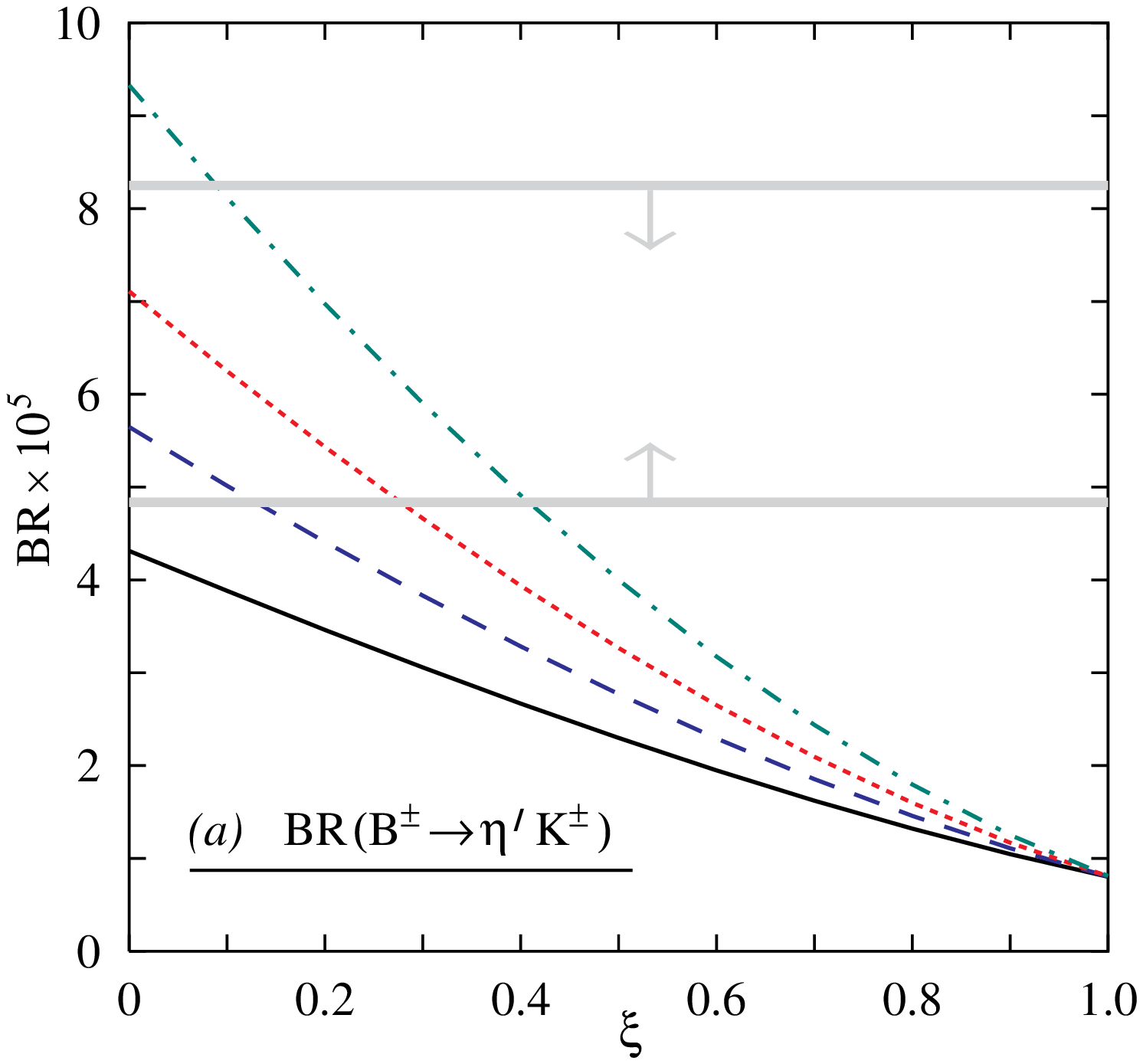}
\hspace*{-1.10cm}
\epsfxsize=6.2cm\epsfysize=7.0cm
\epsfbox{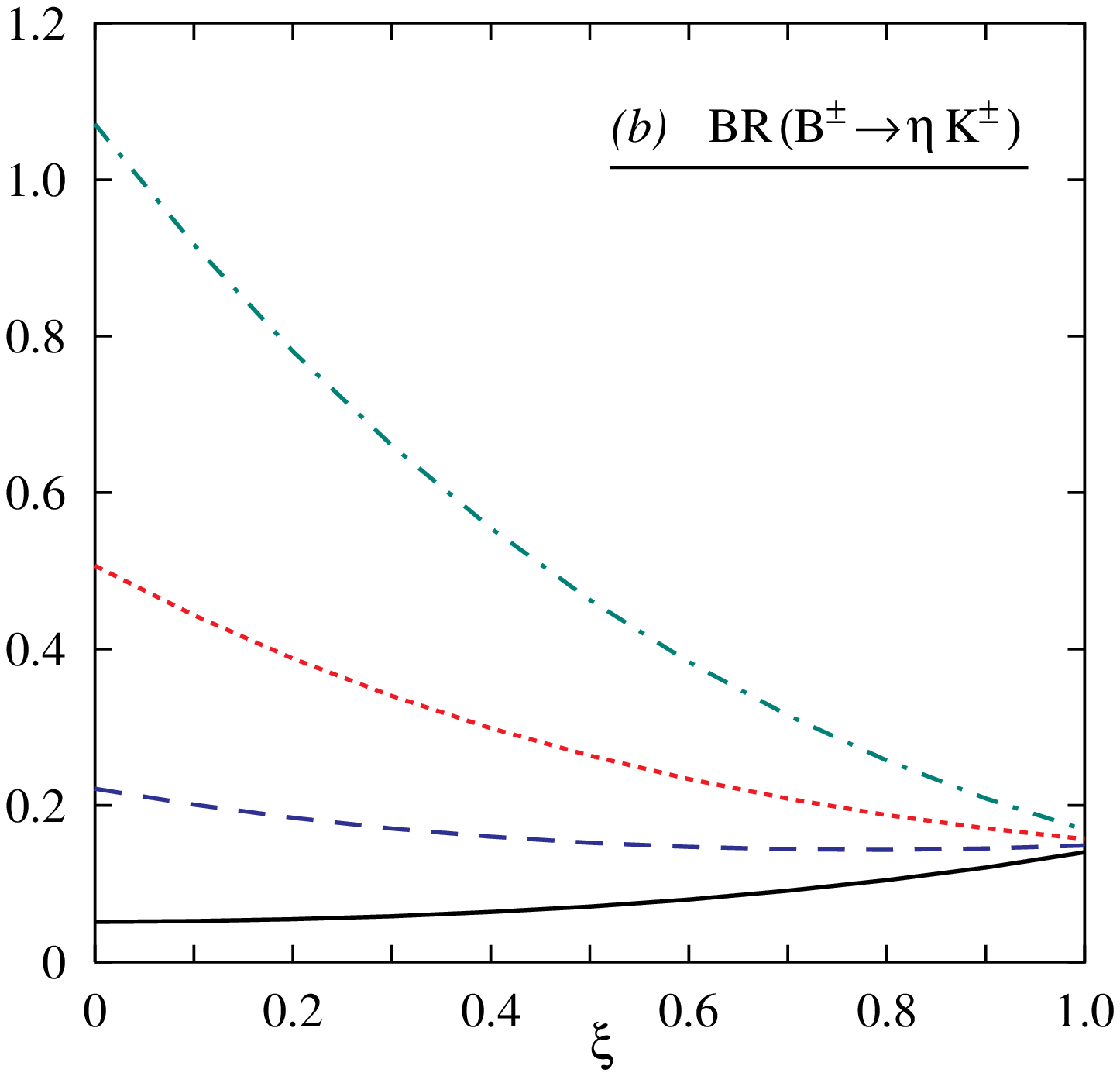}
\hspace*{-1.10cm}
\epsfxsize=6.2cm\epsfysize=7.0cm
\epsfbox{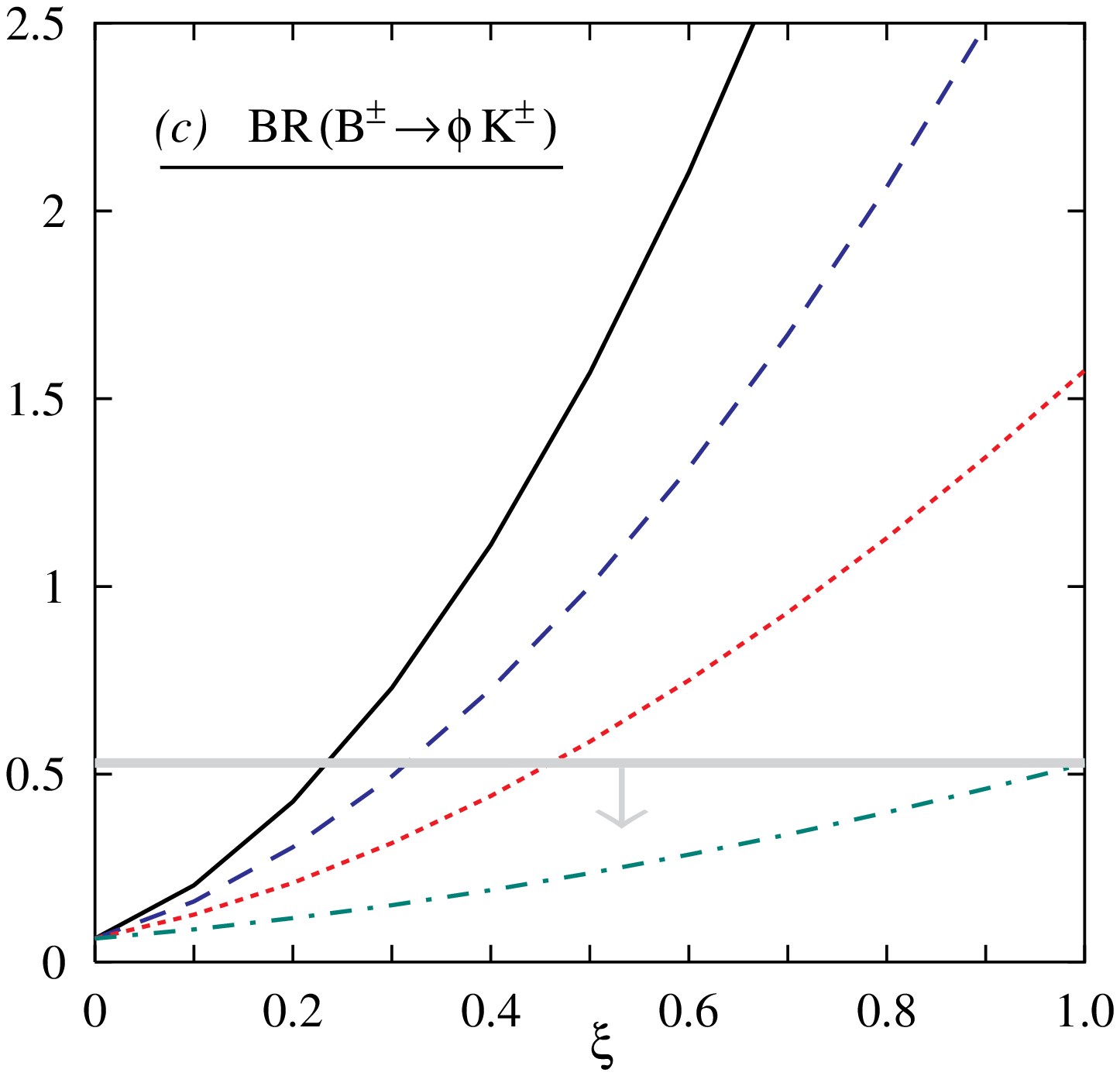}   
\vspace*{-0.6cm}
}
\caption[fig:fig1]{\em Branching ratios for various decays as a function 
       of \ $\xi$. The solid curve gives the SM value.
       In the presence of a $d^R_{222}$ operator with a sfermion mass of 
       200 GeV, 
       the long-dashed, short-dashed and dot-dashed 
       curves correspond to the cases where each of the 
       two $\lambda'$s equal 0.09, 0.07 and 0.05 respectively.
       The thick lines correspond to the experimental bounds.
                   } \label{fig:d_222}
\end{figure}

\begin{figure}[htb]
\centerline{
\epsfxsize=8.5cm\epsfysize=7.5cm
\epsfbox{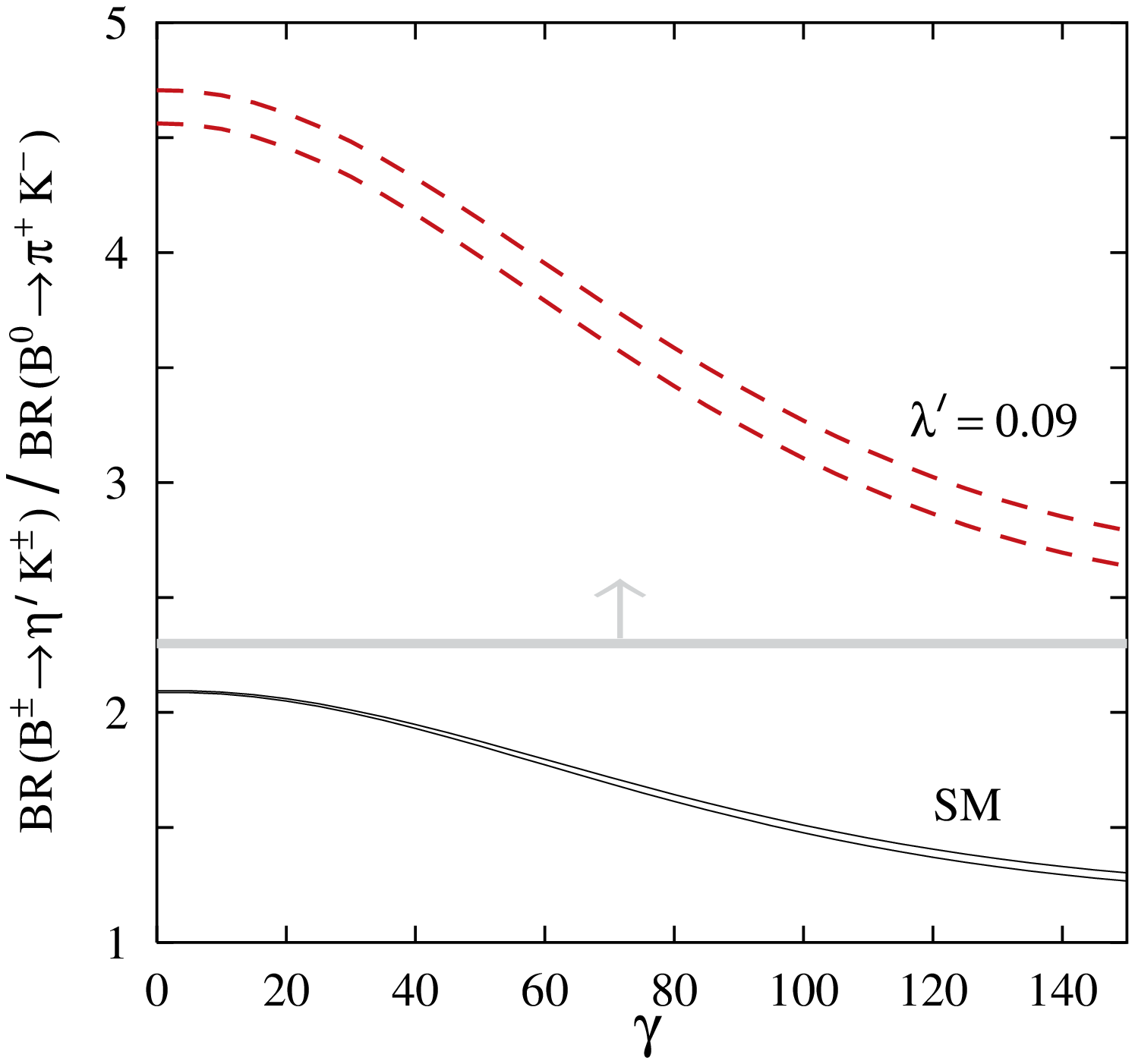}
}
\vspace*{-0.6cm}
\caption[fig:fig2]{\em The ratio $BR(\bpletapk) / BR(B^0 \ra \pi^+ K^-)$
        for $\xi = 0$ as a function of the CKM parameter $\gamma$. The solid 
        curves represent the SM prediction while the dashed 
        curves are for a $d^R_{222}$ with each $\l' = 0.09$. 
        In each case, the upper and lower curves are for 
        $m_s(1 GeV) = 150 (165)$ MeV respectively.
                   } \label{fig:ratio}
\end{figure}

We are now ready to discuss our results. Our goal is to explain 
the branching ratio for the $\bpletapk$ decay while satisfying
the experimental numbers (limits) for all other related decays 
(see Table~\ref{tab:1}). To set the perspective, 
consider the solid curve in 
Fig.~\ref{fig:d_222}($a$), wherein we have plotted 
$BR(\bpletapk)$ as a function of $\xi$. It is quite apparent that 
only for very small $\xi$ could we hope to reconcile the SM predictions
with the observations. One may argue, though, that such a conclusion is 
unwarranted in view of the uncertainty in other parameters such 
as $F$, the CKM elements $V_{cb}$ and $V_{ub}$, the angle $\gamma$ of the
unitarity triangle,
and the strange quark mass \footnote{
     The branching ratio of $\betapk$ increases slightly with
     the increase of the $\eta-\eta'$ mixing angle $\theta$
     \protect\cite{desh}, but since the experimental constraint 
     on this mixing angle is rather tight, we will not consider it
     here.}. 
Consider instead the ratio $BR(\bpletapk)/ BR(B \ra \pi^+ K^0)$ 
which is independent of $F$ and $V_{cb}$. 
In Fig.~\ref{fig:ratio}, we plot this ratio as a function of $\gamma$ 
for $\xi = 0$, so as to maximize it. Clearly, the SM prediction 
falls well below the experimental number (remember that $\gamma \sim 0$
is unable to account for the observed CP violation in $K$-system). 

Thus, if we demand that $\rpv$ solve the $\betapk$ anomaly, the 
relevant  operators need to add {\em constructively} to the SM 
amplitudes. We make a simplifying 
assumption here. Rather than consider the most general case, 
we restrict ourselves to  {\em exactly one non-zero
product} in eq.(\ref{rp_hamilt}) and discuss its consequences. 
This immediately restricts us to particular signs for each of the  
combinations. To wit, we need one of $d^R_{222}$, $d^R_{112}$, $u^R_{112}$ 
and $d^L_{112}$ to be positive. 
On the other hand, only negative values for the other four combinations
$d^L_{222}$, $d^L_{121}$, $d^R_{121}$ and $d''_{112}$ 
could explain the enhanced BR. We shall concentrate on only the first 
set.

It is easy to see that $u^R_{112}$ also enhances the  
$BR(B\rightarrow \pi^+K^-)$. Since there exists a 
stringent experimental bound on this mode, 
the largest allowed value for $u^R_{112}$ 
is too small to explain the $BR(\betapk)$.  
Similarly, the small enhancement due to $d^L_{112}$, which 
occurs only for large value of $\xi$, is unable to explain the anomaly.  
Thus, we are left with only two terms, namely $d^R_{222}$ and $d^R_{112}$.

Let us first focus on $d^R_{222}$. 
In all our subsequent discussions, we take, without any loss of generality,
both the $\l'$s in the product $d^R$ to be equal, and the intermediate 
$i$-th sneutrino mass to be 200 GeV. As is evident from eq. \ref{rp_hamilt},
the new physics contribution is proportional to ${\l'}^2/\msnus$. Since the 
products $\l'_{122}\l'_{123}$ and $\l'_{222}\l'_{223}$ have a stronger 
experimental upper bound than the numbers we need, the only possible  
solution is for $i=3$, {\em i.e.}, $\l'_{322}\l'_{323}$. 
Similar conclusions follow for $d^R_{112}$. 

In Figs. (\ref{fig:d_222}$a$) and (\ref{fig:d_222}$b$), 
we show the effect of a non-zero $d^R_{222}$
on two particular BRs, namely those for
$B^\pm\ra \eta' K^\pm$ and $B^\pm\ra \eta K^\pm$. 
Clearly, a resolution of the anomaly is now possible, 
albeit for a $\l'$-dependent range for $\xi$. 
Since the $\rpv$ contribution to the decay amplitude tends to become
too large with increasing $\l'$, progressively larger values of 
$\xi$ are required. 
As for the other modes, it is easy to see that 
the our solutions respect the experimental numbers/constraints. 
For example, with $\l'=0.09 (0.07)$ and $\xi=0.2(0.3)$, we expect 
$BR (B^0\ra \eta' K^0) = 5(5.5)\times 10^{-5}$, well in consonance 
with observations (Table~\ref{tab:1}). 
Similarly, the BRs for the modes $B^+\ra 
\eta K^{*+}, \eta'K^{*+}$ and $B^0\ra \eta K^0,  \eta K^{*0}, \eta' K^{*0}$ for
$\l'=0.09 $ are predicted to be $1.2(0.6), 0.5(0.3), 0.8(0.4), 0.9(0.4),$ 
and $0.3(0.2)$ ($\times 10^{-5}$) respectively for $\xi=0 (0.5)$.
In fact, if our explanation be the correct one, 
we would expect to see the decay 
$B^\pm\ra \eta K^{*\pm}$ quite soon, whereas some of the other modes  
may be visible in the upcoming B-factories.

At this stage, a comment is in order. For 
Figs.(\ref{fig:d_222}$a,b,c$), we have used 
$F=0.36$ and $m_s (1 GeV)=165$ MeV ($m_s (m_b)=118$ MeV), values preferred by the SM fit.
However, in the presence of additional interactions, 
one may use a different set. As Fig.~\ref{fig:ratio} shows, the 
dependence on $m_s$ is marginal. On the other hand,
a larger value for $F$ would enhance the BRs. 
For example, for $F=0.4$, $\xi=0.55$ and each $\l'=0.09$, the theoretical
BRs for the modes $\eta' K^+$, $\eta' K^0$, $\eta K^+$, $\eta K^0$,
$\pi^-K^+$, $\pi^+ K^0$, $\pi^+\pi^-$ and $\pi^+\pi^0$ (last four modes do not
have any contribution from $d^R_{222}$) are $4.9$, $7.6$,
$0.6$, $0.7$, $2.1$, $2.8$, $1.1$ and $0.9$ respectively (all in units 
of $10^{-5}$). This is the maximum value of $F$ that can be used in
conjunction with $\xi=0.55$ since 
($i$) the prediction for the $\pi^-K^+$ 
mode actually saturates the experimental number, 
 and ($ii$) the data on $B\ra\pi\pi$  implies 
that $F|V_{ub}/V_{cb}|\leq 0.024$ 
(note that semileptonic decays give $|V_{ub}/V_{cb}|=0.08\pm 0.02$). 
Of course, the above does not preclude smaller values for $F$: with 
$F=0.33$, $\xi=0$ and each $\l'=0.09$, the theoretical predictions
for the abovementioned  eight modes are $7.8$, $7.6$, $0.9$, $0.7$,
$1.8$, $2.9$, $1.1$ and $0.8$ respectively (again, all in units of
$10^{-5}$).  Anyway, for these numbers, particularly for those in the 
first set, one can easily see that more and more channels get close
to the discovery limit.

What about the $B\ra PV$ modes? As Fig.~\ref{fig:d_222}($c$) shows,
the SM fit requires $\xi<0.23$. This is
in conflict with other $PV$ modes such as $B^{\pm}\ra \omega K^{\pm}$  
and $B^{\pm}\ra \omega \pi^{\pm}$. The former  requires 
either $\xi<0.05$ or $0.65<\xi<0.85$ while the latter requires
$0.45 < \xi < 0.85$ \cite{desh1}. 
Interestingly, the $d^R_{222}$ operator affects $B^\pm\ra \phi K^\pm$
while the other two decay modes are blind to it. Since this
additional contribution interferes destructively with the SM amplitude, 
$BR(B^\pm\ra \phi K^\pm)$ is suppressed leading to a wider allowed range
for $\xi$ (see Fig~\ref{fig:d_222}$c$). For example, with $\l'=0.09$,
$\xi$ can be as large as $0.8$, thus allowing for a common 
fit to all the three ($PV$) modes under 
discussion\footnote{Note that the favoured value of $\xi$ for the $PP$ 
        and $PV$ modes still continue to be different. While this is {\em not}
        a discrepancy, a common $\xi$ for both these sets can  be 
        accommodated for values of $\l'$ slightly larger than that we 
        have considered.}. 
$d^R_{222}$ also affects a $VV$ decay modes such as 
($B\rightarrow\phi K^*$). As this calculation involves a few more 
model  dependent parameters, 
we do not analyse it here. 

\begin{figure}[htb]
\centerline{
\epsfxsize=8cm\epsfysize=6.9cm
\epsfbox{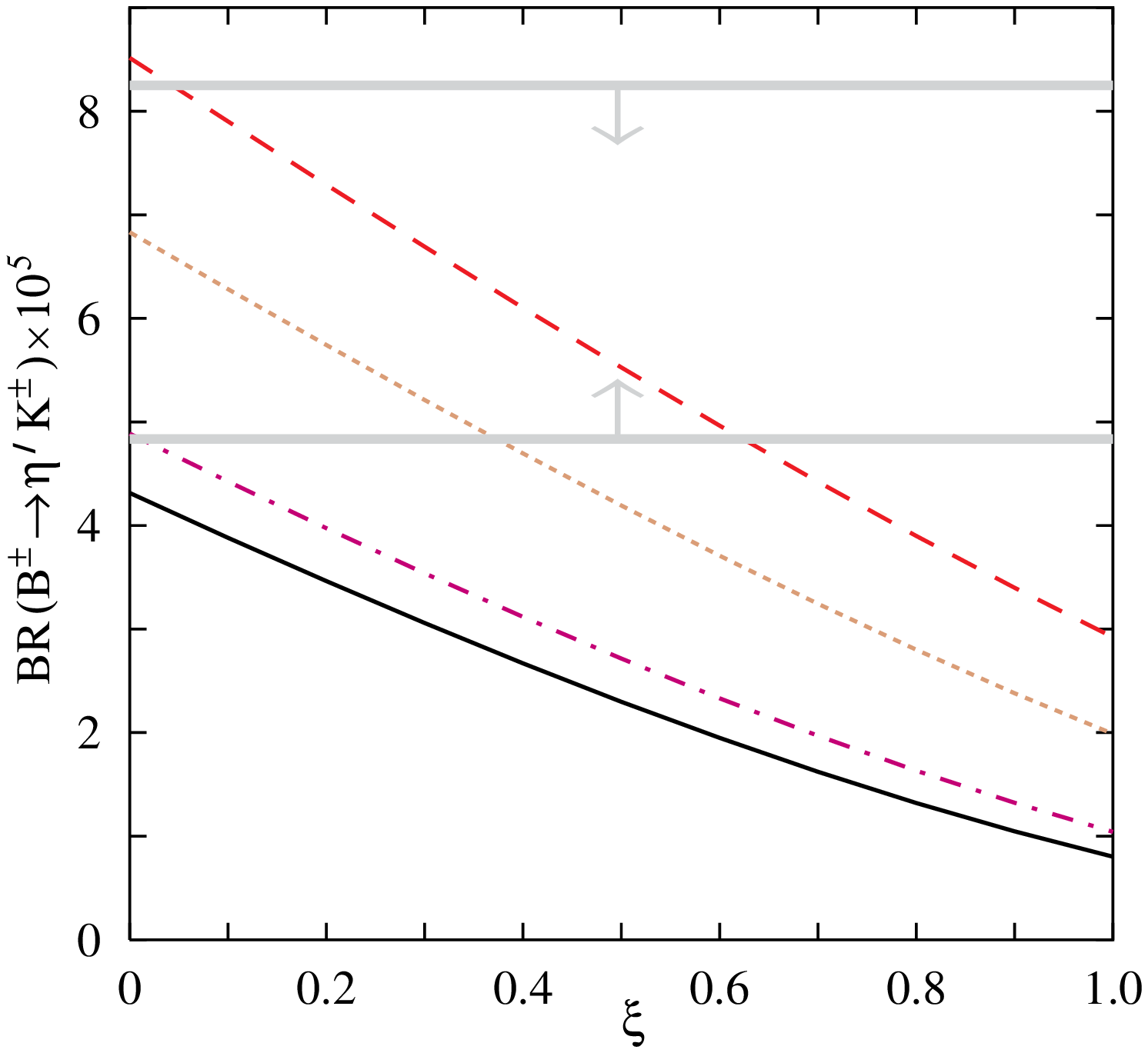}
}
\vspace*{-0.6cm}
\caption[fig:fig1]{\em Branching ratio for  
  $B^{\pm}\ra \eta^{\prime}K^{\pm}$ as a function of \ $\xi$. 
       The solid curve gives the SM value.
       In the presence of a $d^R_{112}$ operator with a sfermion mass of 
       200 GeV, 
       the long-dashed, short-dashed and dot-dashed 
       curves correspond to the cases where each of the 
       two $\lambda'$s equal 0.025, 0.02 and 0.01 respectively.
       The thick lines correspond to the experimental bounds.
                   } \label{fig:d_112}
\end{figure}

Finally, we investigate the consequences for a non-zero $d^R_{112}$
as opposed to $d^R_{222}$. For brevity's sake, we 
present graphs (see Fig.~\ref{fig:d_112}) only for $BR(\bpletapk)$. 
It is interesting to note that $\l'>0.05$ is not admissible
for any $\xi<1$, as the model predictions become significantly 
larger than the observed width. As for $B^0\ra \eta' K^0$, 
the BR is $6.2(4.8)\times 10^{-5}$ for $\xi = 0.3$ and 
$\l' = 0.025 (0.02)$ (see Table~\ref{tab:1}).  
Indeed, the entire parameter space allowed by $B^\pm\ra \eta' K^\pm$ 
is also allowed by $B^0\ra \eta' K^0$. 
%
%
For such values of $\l'$s, $BR(B^\pm\ra \eta K^\pm \lappeq 3\times 10^{-6}$,
and thus well below the experimental upper limit. 
Similarly, for the other relevant $PP$ modes 
$B^0\ra \eta K^0, \pi^0K^0$ and $B^\pm\ra K^\pm\pi^0$, the
maximum BRs are $0.15$, $1.9$ and $1.4$ ($\times 10^{-5}$) respectively. 
Since, for all these decays, the $\rpv$ contribution interferes 
destructively with the SM one, the resultant predictions are 
considerably suppressed. The best constraints emanate from 
$BR(B^\pm\ra K^0\pi^\pm)$ which supports $0.03<\xi<0.8$ for the 
$\l'$s used in Fig.~\ref{fig:d_112}.

The case for the  $PV$ modes is similar. 
For the decays $B^+\ra \eta K^{*\pm}, \eta' K^{*\pm},
\pi^0 K^{*\pm}$ and $B^0\ra \eta K^{*0},\eta' K^{*0}, \pi^0 K^{*0}$  the $\rpv$
operator adds constructively whereas for $B^\pm\ra K^0\rho^\pm,
\pi^\pm K^{*0}$ the interference is destructive in nature. 
The maximum possible  
BRs for the first six modes, for $\l'=0.03$ and $\xi=0(0.5)$, are
$1.2(1.0), 0.75(0.45), 0.4(0.3), 1.7(1.1), 1.0(0.5), 0.8(0.4)$ 
($\times 10^{-5}$) respectively, smaller than the corresponding experimental 
numbers. For the last two modes, of course, no question of contradiction with
experiment arises. 

In short, the modes  $B^\pm\ra \eta K^{*\pm}$ and $B^0\ra \eta K^{*0}$ 
are close to the discovery limit whereas other modes may have to wait 
for the next generation B-machines. 
In a subsequent paper\cite{us}, we will discuss the 
CP violating effect of these $\rpv$ operators  on all
these, and other, modes in detail.

\vskip 1 true cm
\centerline{\bf 5. Conclusion}

To conclude, we have written down all possible $\rpv$ SUSY contributions  
to the
effective Hamiltonian for the $B^{\pm}\ra \eta' K^{\pm}$ decay. We have 
found that only two new terms, each involving two 
$\l'$-type couplings, can raise the BR to satisfy the 
experimental number. We have shown that though these two terms 
appear in other nonleptonic decay modes of the $B$
meson, their BRs always satisfy the experimental constraints in the whole 
of the allowed parameter space of $\l'$,
$\msnu$ and $\xi$. Modes like $\eta K^{*+}$, $\eta K^{*0}$ are close to their
discovery limits. Further, one of the new contributions allows 
larger parameter 
space in $\xi$ for the decay $B^{\pm}\ra \phi K^{\pm}$,   
where the other observed modes {\em e.g.},
$B^{\pm}\ra \omega K^{\pm}$ and $B^{\pm}\ra \omega \pi^{\pm}$ can be fit; 
this is not possible in the SM framework.
This leads us to believe that $B$-decays and upcoming 
B-factories may be the most
promising place to look for new physics beyond the SM.

\vspace*{3ex} 
We thank  Amitava Datta and N.G. Deshpande for illuminating discussions.

\end{document}